\documentclass[twocolumn,aps,prl,preprintnumbers,floatfix]{revtex4}
\usepackage{bm}
\usepackage{amsmath}
\usepackage{graphicx}
\usepackage{color}
\usepackage{hyperref}
\hypersetup{
     unicode=false,          
     pdftoolbar=true,        
     pdfmenubar=true,        
     pdffitwindow=false,     
     pdfstartview={FitH},    
     pdftitle={My title},    
     pdfauthor={Author},     
     pdfsubject={Subject},   
     pdfcreator={Creator},   
     pdfproducer={Producer}, 
     pdfkeywords={keyword1} {key2} {key3}, 
     pdfnewwindow=true,      
     colorlinks=false,       
     linkcolor=red,          
     citecolor=green,        
     filecolor=magenta,      
     urlcolor=cyan           
}

\begin{document}

\title{Non-Quantized Edge Channel Conductance and Zero 
Conductance Fluctuation in Non-Hermitian Chern Insulators}
\author{C. Wang}
\email[Corresponding author: ]{cwangad@connect.ust.hk}
\affiliation{School of Electronic Science and Engineering
and State Key Laboratory of Electronic Thin Film and
Integrated Devices, University of Electronic Science 
and Technology of China, Chengdu 610054, China}
\author{X. R. Wang}
\email[Corresponding author: ]{phxwan@ust.hk}
\affiliation{Physics Department, The Hong Kong University of 
Science and Technology, Clear Water Bay, Kowloon, Hong Kong}
\affiliation{HKUST Shenzhen Research Institute, Shenzhen 518057, China}
\date{\today}

\begin{abstract}
The quantized conductance plateaus and zero conductance fluctuation 
are the general consequence of electron transport in chiral edge 
states of Hermitian Chern insulators. Here we show that the 
physics of electron transport through chiral or helical edge 
channels becomes much richer when the non-Hermicity is allowed. 
In the presence of an unbalanced non-Hermicity where the 
chiral edge states have finite-lifetime, the conductance of 
the edge channels is not quantized, but its 
conductance fluctuation is zero. 
For the balanced non-Hermicity, however, the chiral edge states 
have infinite-lifetime, and the conductance is quantized as in 
the case of Hermitian Chern insulators. Both non-quantized and 
quantized conductance plateaus of zero conductance fluctuation 
are robust against disorders. We present a theory that explains 
the origin of the non-quantized 
conductance plateaus. The physics revealed here should also be 
true for the chiral and helical surface states in other topological 
materials such as quantum anomalous Hall systems, 
Weyl semimetals, and topological superconductors.
\end{abstract}
\maketitle

\emph{Introduction.}$-$Searching various topological insulators 
\cite{Kane1,Kane2,Bernevig,Konig,Roth,Hasan,Qi,Moore,Chang}, topological
semimetals \cite{Wan,Yang,Burkov,Weng,Huang,Xu,Lv,Lu,Shekhar,Burkov2}, 
and topological superconductors \cite{Fu,Qi2,Potter,Nandkishore,Grover,Xu2,He} 
has attracted enormous attention in recent years because of 
their fundamental interest and possible applications of the 
exotic properties of topologically-protected surface states. 
The topological states can exist not only in the electronic systems, 
but also in other classical or quantum systems such as electromagnetic 
waves \cite{Haldane,Wang,Khanikaev,Sun,Ozawa}, mechanical waves 
\cite{Kane3}, and spin waves \cite{Fujimoto,Katsura,Onose,Matsumoto,
Mena,Lee,Chisnell,Kim,Wangxs,Su2}. The topological band theory 
\cite{Bansil,Chiu,Armitage} is believed to play a crucial role in 
understanding the physics of these newly found quantum phases. 
As a well-accepted paradigm, the quantized conductance $e^2/h$ 
observed in magnetic doped Chern insulators (CIs), known as the 
quantum anomalous Hall effect, is the consequence of the 
impurity-immune chiral edge states \cite{Sheng,Fulga} in  
Hermitian CIs. 
\par

The topological band theory is based on the Hermitian Hamiltonian 
for closed quantum systems with conserved particle probability. 
Strictly speaking, all interesting systems that one wants to 
understand and to probe are not closed. The Hermicity of an 
open system is universally lost, and the system always involves 
certain degrees of gain and loss. For example, the inevitable 
electron-electron, electron-impurity, and electron-photon 
scatterings in an electronic system lead to complex self-energies  
\cite{Kozii,Papaj,Shen,Zyuzin} of single electron states such 
that the lifetime of a single electron state is always finite. 
Recently, motivated by this consideration, the non-Hermitian 
topological nontrivial systems have been intensively investigated \cite
{Esaki,Lee1,Liang,Leykam,Menke,Li,Xiong,Alvarez,Gong,Harari,Xu3,Ni,Yanghh}.
A generalization of the topological band theory to the non-Hermitian
systems has been done, which highlights the breakdown of 
the general bulk-boundary correspondence \cite{Shen2,Yao1,Yao2}. 
While most of the theoretical studies focused on static properties 
such as particle spectrum and wavefunctions, the transport 
properties of the non-Hermitian topological states are less studied,
although they are directly measured in experiments.
\par

In this letter, we compute the conductance of a piece of 
non-Hermitian CI Hall bar. 
In the case of a balanced non-Hermicity, an electron injected into a 
chiral edge channel from one electrode will eventually reach the other 
electrode without losing its quantum coherence, and the conductance is quantized to the integer of $e^2/h$, 
the same as that of the Hermitian CIs, and robust against disorders. 
The quantized conductance disappears and diminishes only in very strong 
disorders when the edge channels are completely destroyed by the Anderson localizations. 
More interestingly, the conductance of systems with an unbalanced 
non-Hermicity is non-quantized due to the finite-lifetime of a 
single electron state. In this situation, conductances 
vary continuously with the non-Hermicity and the Hall bar length. 
We call this phase the anomalous Chern insulators. By adding weak 
on-site disorders, the anomalous CI phase is then identified by the 
non-quantized conductance plateaus with zero conductance fluctuation. 
We develop an effective theory to explain the appearance of non-quantized 
conductance plateaus. The theory explains perfectly the simulation results. 
\par  

\emph{Non-Hermitian Hamiltonian.}$-$We consider the following 
two-dimensional (2D) tight-binding Hamiltonian on a square lattice,  
\begin{equation}
\begin{gathered}
H=\sum_{\bm{i}} \left[c^\dagger_{\bm{i}}\left(m\sigma_z+i\vec{\gamma}\cdot
\vec{\sigma}+V^1_{\bm{i}}\sigma_0+V^2_{\bm{i}}\sigma_z\right)c_{\bm{i}}\right]\\
-\sum_{\bm{i}} \dfrac{t}{2}\left[c^\dagger_{\bm{i}+\hat{x}}\left(\sigma_z+
i\sigma_x\right)c_{\bm{i}}+c^\dagger_{\bm{i}+\hat{y}}\left(\sigma_z+
i\sigma_y\right)c_{\bm{i}}+H.c.\right],
\end{gathered}\label{hamiltonian}
\end{equation}
where $c^{\dagger}_{\bm{i}}=(c^{\dagger}_{\bm{i},\uparrow},c^{\dagger}_{\bm{i},\downarrow})$ 
and $c_{\bm{i}}$ are the single electron creation and 
annihilation operators on lattice site $\bm{i}=(x_i a,y_i a)$ with 
$x_i$ and $y_i$ being integers and $a$ the lattice constant. 
$m$ and $t>0$ are respectively the Dirac mass and the hopping energy. 
$\vec{\sigma}=(\sigma_x,\sigma_y,\sigma_z,\sigma_0)$ 
with $\sigma_{x,y,z}$ being Pauli matrices and $\sigma_0$ 
identity matrix, acting on spin (or pseudospin) space.
Hamitonian \eqref{hamiltonian} is Hermitian in the absence of 
$\gamma$-terms, $i\vec{\gamma}\cdot\vec{\sigma}$, where 
$\vec{\gamma}=(\gamma_x,\gamma_y,\gamma_z,\gamma_0)$ with 
$\gamma_{x,y,z,0}/t$ being small real numbers that measure the 
degrees of non-Hermicity. This form of non-Hermicity has 
been widely used in literature 
\cite{Alvarez,Esaki,Lee,Menke,Ni,Xiong,Xu3,Yao1,Yao2,Shen2} 
and was derived from the electron-electron, electron-photon, and 
electron-impurity interactions \cite{Kozii,Papaj,Shen,Zyuzin}.
Disorders are introduced through $V^1_{\bm{i}}/t$ and $V^2_{\bm{i}}/t$ 
that distribute randomly and uniformly in the range of $[-W/2,W/2]$.
\par

In the clean limit of $W=0$, Eq.~\eqref{hamiltonian} can be 
block-diagonalized in the momentum space as  
$H=\sum_{\bm{k}}c^{\dagger}_{\bm{k}}h(\bm{k})c_{\bm{k}}$ with 
\begin{equation}
\begin{gathered}
h(\bm{k})=
(t\sin k_x+i\gamma_x)\sigma_x
+(t\sin k_y+i\gamma_y)\sigma_y\\
+(m-t\cos k_x-t\cos k_y+i\gamma_z)\sigma_z+i\gamma_0\sigma_0.
\end{gathered}\label{hamiltonian2}
\end{equation}
The Hermitian part of Eq.~\eqref{hamiltonian2} is the Qi-Wu-Zhang model 
\cite{QWZ}. For $0<|m|<2t$, this model supports the quantum anomalous 
Hall phase with chiral edge states that form a chiral edge channel. 
The non-Hermicity $i\gamma_j\sigma_j$ (here $j=x,y$), acting on the 
eigenstates of the Hermitian part, adds a pre-factor 
$\exp[\gamma_j x_j/t]$ to the Bloch wavefunction of the Qi-Wu-Zhang model.
(This can be seen either from the differential equation form 
of the continuum model of \eqref{hamiltonian2} near $\bm{k}=0$ 
by replacing $k_j$ with $-i\partial/\partial x_j$ or from the 
perturbation theory by treating the non-Hermitian part as the 
perturbation.) 
Thus, for finite-size systems with the open 
boundary condition, the non-Hermitian potential 
causes the Bloch states with well-defined momentum to exponentially 
localize at sample boundaries. This notable phenomenology is 
referred as the non-Hermitian skin effect \cite{Yao1,Yao2}.     
\par

\begin{figure}[ht!]
\centering
\includegraphics[width=0.45\textwidth]{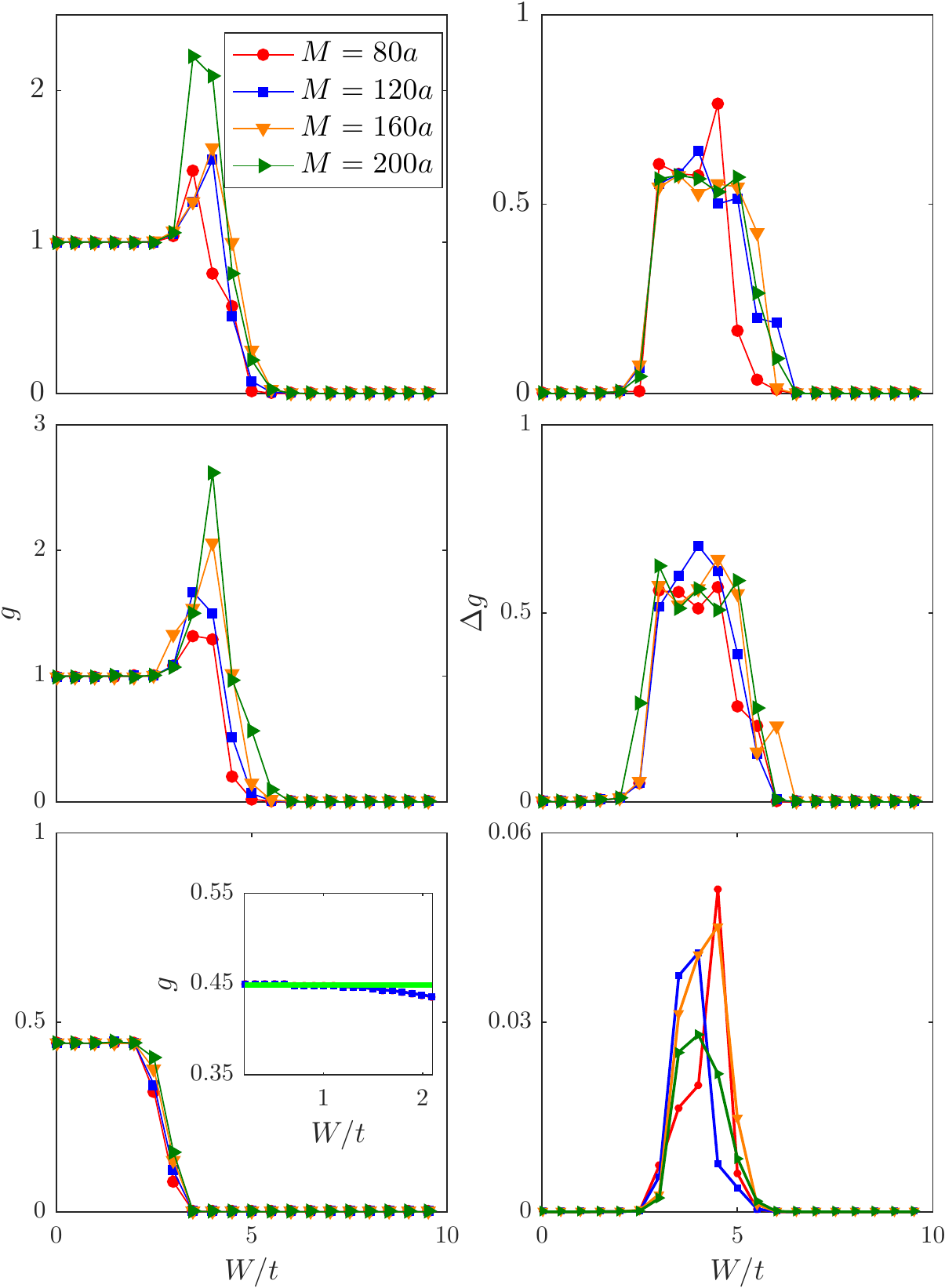}
\caption{(a,b) $g$ (a) and $\Delta g$ (b) vs $W$ for 
$\vec{\gamma}_y=(0,\gamma,0,0)$. 
(c,d) Same as (a,b) for $\vec{\gamma}_z=(0,0,\gamma,0)$. 
(e,f) Same as (a,b) for $\vec{\gamma}_0=(0,0,0,\gamma)$. 
The inset in (e) is an enlargement of the non-quantized
conductance plateau. Green line is $g=0.449$.
Here $\gamma=10^{-3}t$, $L=400a$, and $M/a=80,120,160,200$.
}
\label{fig1}
\end{figure}  

\emph{Anomalous Chern insulator.}$-$To reveal the electron 
transport properties of non-Hermitian CIs, we calculate 
the conductance of a rectangle Hall bar of length $L$ and 
width $M$ connected to two semi-infinite electrodes  
at two ends along the $x$ direction. Electrons in the Hall 
bar are govern by Eq.~\eqref{hamiltonian} while electrons  
in the leads are assumed to be normal free Fermi gas. 
Thus, the non-Hermicity exists only in the disordered Hall 
bar and leads are Hermitian. The conductance $G$ of the 
two-terminal setup is numerically computed by the 
Landauer-B\"{u}ttiker formalism \cite{Landauer,Buttiker}.
Without losing generality, we consider three different 
non-Hermicities: $\vec{\gamma}_y=(0,\gamma,0,0)$,  
$\vec{\gamma}_z=(0,0,\gamma,0)$, and 
$\vec{\gamma}_0=(0,0,0,\gamma)$ \cite{gammax}.
We will show below that the first two non-Hermicities are 
fundamentally different from that of the last case. 
\par

The dimensionless conductance $g=\langle G\rangle/(e^2/h)$ and 
the conductance fluctuation $\Delta g=\sqrt{\langle G^2\rangle-
\langle G\rangle^2}/(e^2/h)$, where $\langle\cdots\rangle$ 
denotes the ensemble average, are calculated by varying 
the disorder strength $W$ and the system width $M$. 
We fix the Dirac mass $m=t$ and the Fermi energy $E=-0.01t$ 
to focus on the edge channel transport \cite{topology}. 
Figure~\ref{fig1} shows $g$ (each point is averaged over more 
than 6000 different configurations) and $\Delta g$ as a 
function of $W$ for $\vec{\gamma}_y$, $\vec{\gamma}_z$, and 
$\vec{\gamma}_0$ with $\gamma=10^{-3}t$ and $L=400a$. 
For the first two cases ($\vec{\gamma}_y$ and $\vec{\gamma}_z$), we 
observed the width-independent quantized plateaus exactly at $g=1$ 
and the vanishing conductance fluctuations $\Delta g=0$ as long as 
$W$ is below a critical value $W_{c}$, see Figs.~\ref{fig1}(a-d). 
The calculated results of the non-Hermitian CIs thus accord quantitatively
with the expectation for Hermitian CIs: $g$ is 
quantized at $1$ within a certain range of disorder ($W<W_c$). 
Beyond $W_c$ ($W>W_c$), $g$ is not quantized and $\Delta g$ is 
non-zero until all conducting channels disappear and $g$ becomes zero. 
The quantized conductance plateau is a direct consequence of the 
existence of the backscattering-immune edge channels. 
\par

Another more striking result is the non-quantized conductance 
plateaus (insensitive to disorder) in the case of $\vec{\gamma}_0$. 
As a representative example, Fig.~\ref{fig1}(e) shows a non-quantized 
conductance plateau of $g=0.449$ for $\gamma=10^{-3}t$ and various 
$M$ range from $80a$ to $200a$. For $W<W_c=t$, the non-quantized 
conductance plateaus do not show any conductance fluctuation, as 
shown in Fig.~\ref{fig1}(f). Interestingly, the value of the 
non-quantized conductance depends neither on the width of the bar nor 
on disorder $W<W_c=t$, a behavior which, together with the emergence of 
the conductance plateaus, is reminiscent of the chiral edge channel 
transport in Hermitian CIs. However, one cannot simply tie our results 
to the non-Hermitian topological band theory \cite{Shen2,Yao1,Yao2} 
that claims generalized Chern numbers for non-Hermitian CIs, in 
contrast to non-quantized $g$ plateaus observed here. 
We thus name the state as the {\it anomalous Chern insulators} that can 
have an arbitrary value of conductance with zero conductance fluctuation. 
This is highly non-trivial because the fluctuation is an intrinsic nature 
of quantum physics under the orthodox statistical interpretation 
of quantum mechanics. The only case of a zero conductance fluctuation 
is the single 1D-unidirectional channel including the quantum Hall 
systems that ties conductance quantization with zero conductance 
fluctuation.
\par

\begin{figure}[ht!]
\centering
\includegraphics[width=0.45\textwidth]{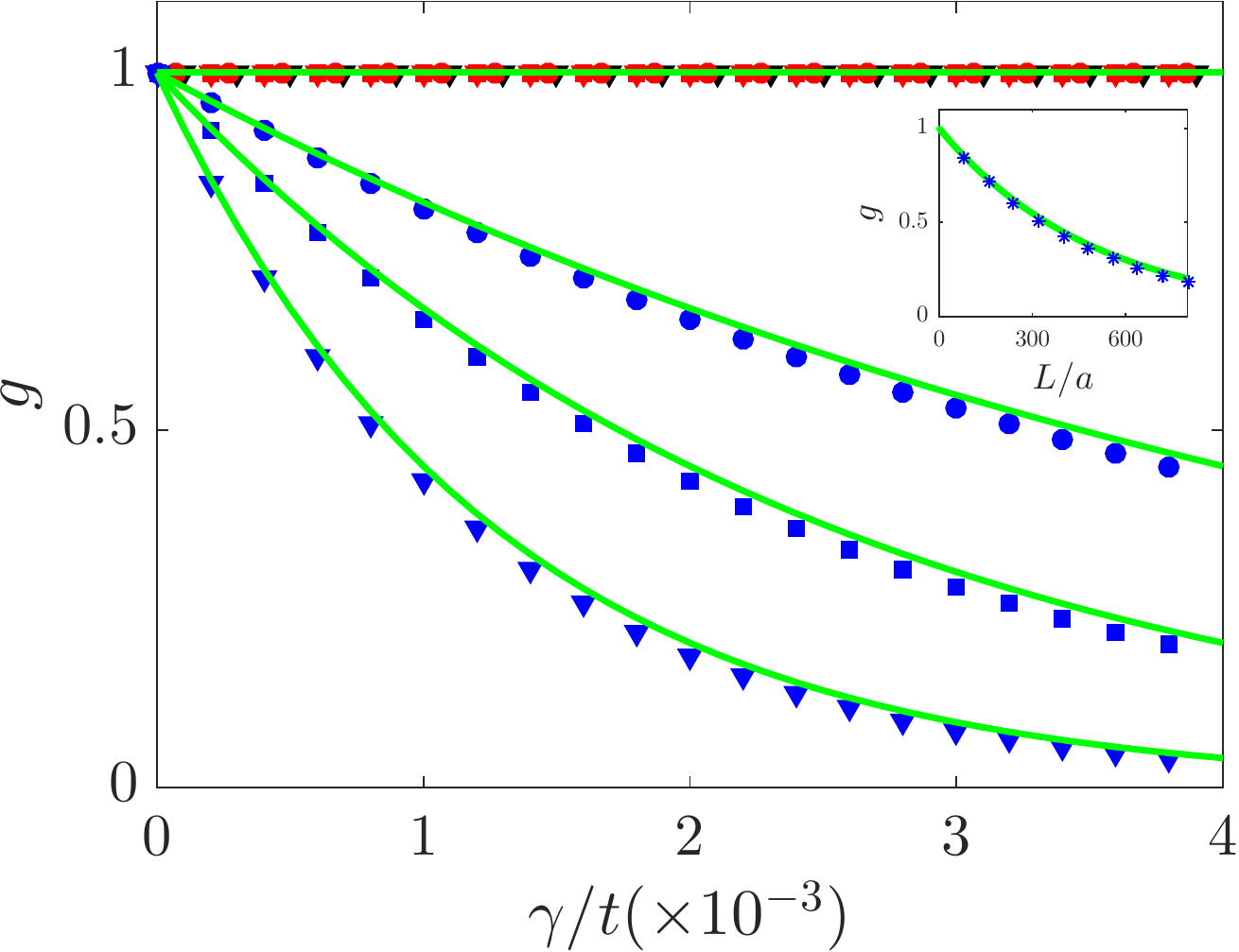}
\caption{$g$ vs $\gamma$ for $\vec{\gamma}_y$ (black), 
$\vec{\gamma}_z$ (red), and $\vec{\gamma}_0$ (blue). 
For each non-Hermicity, the system lengths are $L/a=400$ 
(triangles), 200 (squares), and 100 (circles). 
Green lines are Eq.~\eqref{spectrum3} without any
fitting parameter. 
Inset: $g$ (blue stars from numerical calculations and 
the green line is Eq.~\eqref{spectrum3}) as a function
of $L$ for $\vec{\gamma}_0$ with $\gamma=10^{-3}t$.
Here $M=80a$ and $W=0.5t$.
}
\label{fig2}
\end{figure}

To further reveal the nature of this anomalous CI phase, we 
investigate how $g$ depends on the bar length $L$ and the 
non-Hermicity strength $\gamma$. Figure~\ref{fig2} shows 
$g(\gamma,L)$ for $\vec{\gamma}_y$, $\vec{\gamma}_z$, and 
$\vec{\gamma}_0$ at a fixed randomness $W=0.5t<W_c$ and sample 
width $M=80a$. Clearly, the quantized conductance plateaus 
is independent to $L$ and $\gamma$ for the cases of 
$\vec{\gamma}_y$ and $\vec{\gamma}_z$. However, for the cases 
of $\vec{\gamma}_0$ where the anomalous CI phase is observed, an 
exponential decay of the conductance with both $L$ and $\gamma$, 
$g=\exp[-c\gamma L]$, is observed for either fixed $\gamma$ 
or fixed $L$ as shown in Fig.~\ref{fig2} and its inset. 
Here $c$ is a constant derived later. Again, zero conductance 
fluctuations always accompany with the conductance plateaus, 
see the Supplemental Materials \cite{supp}.      
\par

\begin{figure}[ht!]
\centering
\includegraphics[width=0.45\textwidth]{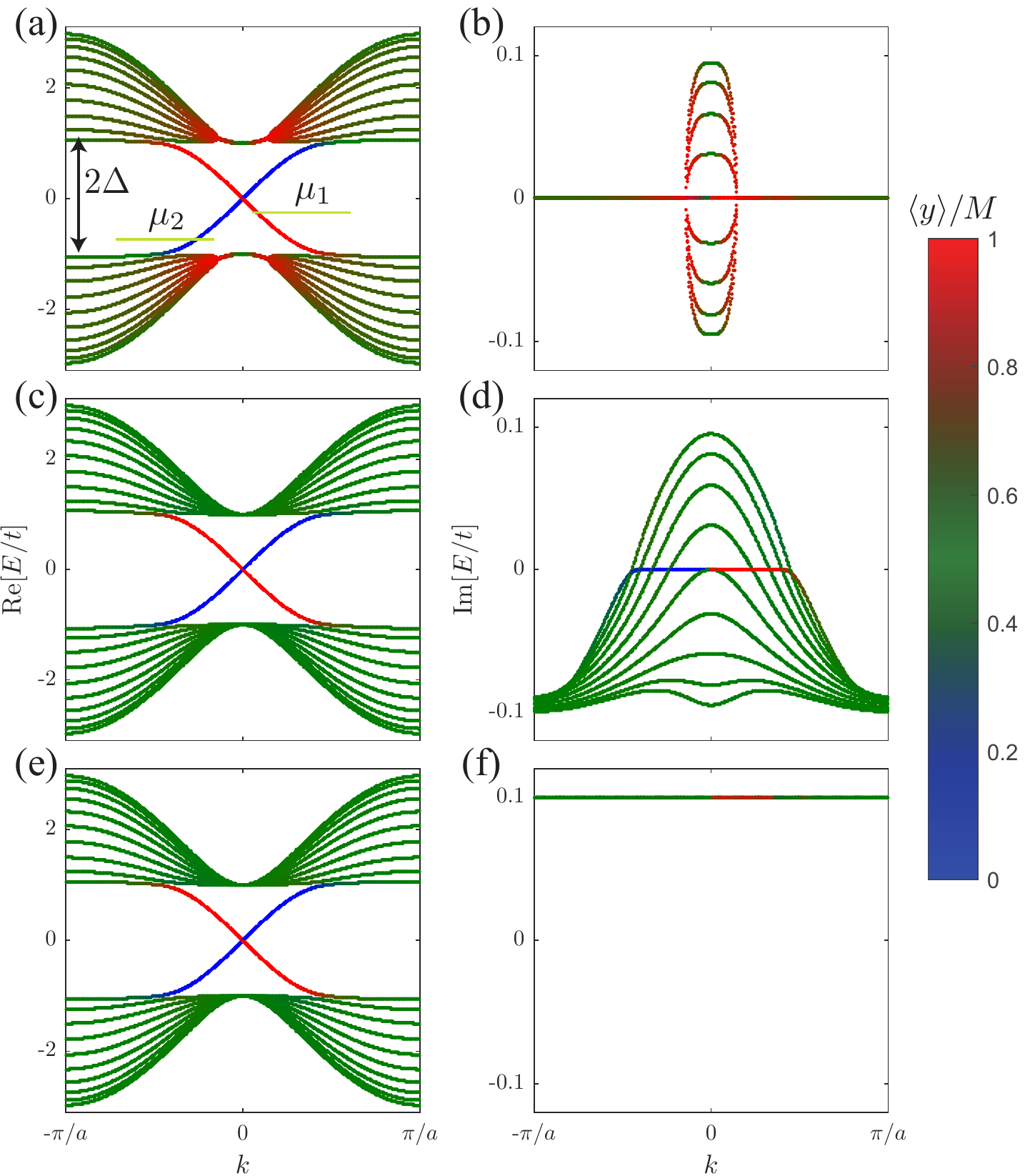}
\caption{Bulk spectra of the clean Hall bar of width $M=20a$.
The chemical potentials of the $+k$ electrons and $-k$ electrons are 
$\mu_1$ and $\mu_2$, respectively. 
(a) $\text{Re}[E/t]$ for $\vec{\gamma}_y$ with $\gamma=0.1t$. 
(b) $\text{Im}[E/t]$ for $\text{Re}[E]<0$ with the same parameters as (a). 
(c,d) Same as (a,b) but for $\vec{\gamma}_z$. (e,f) Same as (a,b) but 
for $\vec{\gamma}_0$. Colors encode $\langle y\rangle/M$.   }
\label{fig3}
\end{figure}

\emph{Origins of anomalous CI.}$-$One crucial question arises 
immediately: Why are the non-quantized conductance plateaus 
of zero fluctuation possible in the anomalous CI phase? From the standard 
topological band theory \cite{Qi}, the chiral edge state of 
Hamiltonian \eqref{hamiltonian2} is the linear superposition 
of both spin-up and spin-down orbits with the equal weight. 
For the quantized conductance with the non-Hermicity of 
$\vec{\gamma}_y$ or $\vec{\gamma}_z$, the gain and loss of the 
two obits compensate with each other. Thus, the local electron 
density in the chiral edge channel is over all conserved  
(real eigenenergies in the edge channel discussed below
and Fig.~\ref{fig3}).  
As a result, each edge channel contributes one conductance 
quanta, leading to a quantized conductance plateau as a Hermitian 
edge channel does. For the case of $\vec{\gamma}_0$, both spin-up 
and spin-down orbits decay, thus an electron in the chiral channel 
has a finite-lifetime (complex eigenenergies in the channel), and 
electron density in the edge channel decays, leading to a non-quantized 
conductance plateau whose value decreases exponentially with the 
length of the Hall bar, but does not depend on the bar width. 
\par   

To obtain an explicit expression that can be compared with the
numerical results, we use the linear response theory \cite{Datta} 
to derive the electron conductance in a decayed chiral edge channel.
In the two-terminal configuration without disorder, the current 
$I_{n,k}$ of a transverse mode (label by $n$) with momentum $k_x=k$ 
is given by
\begin{equation}
\begin{gathered}
I_{n,k}=\rho_{n,k} v^{g}_{n,k}f(\text{Re}\left[E_{n}\right])T(E_n),
\end{gathered}\label{current1}
\end{equation}
where $\rho_{n,k}$ is the electron charge density, $T(E_n)$ is the 
transmission coefficient of state $E_n$, $f$ and $v$ are respectively
the Fermi-Dirac distribution and the group velocity. 
$E_n(k)$ is the energy dispersion relation of the transverse mode. 
We assume the reflectionless contacts so that $+k$ electrons in the edge 
channel come from the left lead and the $-k$ electrons from the right 
lead with chemical potentials of $\mu_1$ and $\mu_2$, respectively. 
In the linear response region, we set $|\mu_2|,|\mu_1|<\Delta$ so that 
no bulk states contribute to the conductance, 
where $2\Delta$ is the bulk gap, as shown in Fig.~\ref{fig3}(a). 
At the zero temperature, $f_{1,2}(\text{Re}\left[E_{n}\right])=\theta
(\mu_{1,2}-\text{Re}\left[E_{n}\right])$ for the $+k$ and the $-k$ states, 
respectively. Here $\theta(x)$ is the Heaviside step function. 
\par 

Since the non-Hermitian Hamiltonian~\eqref{hamiltonian} 
breaks the parity-time-reversal symmetry \cite{Hu}, the energies of 
eigen-modes are complex, see Fig.~\ref{fig3}. 
The dispersion relation of the real part relates to the electron group 
velocity and imaginary part is the lifetime of the electrons in the 
modes so that the particle density in the non-Hermitian CIs is 
\begin{equation}
\begin{gathered}
\rho_{n,k}=\dfrac{e}{L}\exp\left[-\dfrac{4\pi\text{Im}\left[E_n\right]}
{h}\tau_{n,k}\right], 
\end{gathered}\label{current2}
\end{equation}
where $e$ and $h$ are respectively the electron charge and the Plank constant. 
$\tau_{n,k}=L/v^{g}_{n,k}$ is the travel time for an electron through 
the Hall bar from one lead to the other. $h/(4\pi \text{Im}
\left[E_n\right])$ is the single electron lifetime in state $(n,k)$.
In general, the non-Hermicity adds a correction to the Hellmann-Feynman 
theorem such that the group velocity differs from that of the 
Hermitian system as discussed in References \cite{Hu1,Schomerus}. 
However, for the constant non-Hermicity considered here, the group 
velocity is  
\begin{equation}
\begin{gathered}
v^{g}_{n,k}=\dfrac{2\pi}{h}\dfrac{d \text{Re}\left[E_n\right]}{d k},
\end{gathered}\label{current3}
\end{equation}
providing that $\gamma\ll t$ \cite{supp}. 
\par

Using Eqs.~\eqref{current1}, \eqref{current2}, and \eqref{current3},
the net current flowing from one lead to the other is \cite{supp}
\begin{equation}
\begin{gathered}
I=\dfrac{e}{h}(e^{\left[-2\text{Im}\left[\epsilon_+\right]L/(ta)\right]}\mu_1-
e^{\left[-2\text{Im}\left[\epsilon_-\right]L/(ta)\right]}\mu_2)\\
+(I^{+}_{\text{bulk}}-I^{-}_{\text{bulk}}).
\end{gathered}\label{current4}
\end{equation}
The first term is the contribution from the edge channels where $\epsilon_{\pm}$ 
denote the edge-state energy for $k>0$ and $k<0$, respectively. 
The second term is the currents to the right (+) and to the left (-) 
due to bulk states, 
\begin{equation}
\begin{gathered}
I^{+(-)}_{\text{bulk}}=\\
\sum_{\text{bulk}}\sum_{k>0(k<0)}
\dfrac{2\pi e}{h L}
\exp\left[-\dfrac{4\pi\text{Im}\left[E_n\right]L}{h v^g_{nk}}\right]
\dfrac{d\text{Re}\left[E_n\right]}{dk}T(E_n).
\end{gathered}\label{current5}
\end{equation}
\par

The dimensionless conductances $g$ can then be derived from 
Eq.~\eqref{current4} for both balanced non-Hermicities of 
$\vec{\gamma}_y$ and $\vec{\gamma}_z$ and unbalanced 
non-Hermicity of $\vec{\gamma}_0$.
\par
For balanced non-Hermicities, as shown in Figs.~\ref{fig3}(a-d), 
both $\text{Re}[E_n]$ and $\text{Im}[E_n]$ of bulk eigenenergies 
are the even functions of $k$,
\begin{equation}
\begin{gathered}
\text{Re}[E_n(k)]=\text{Re}[E_n(-k)],
\end{gathered}\label{spectrum1}
\end{equation}
and
\begin{equation}
\begin{gathered}
\text{Im}[E_n(k)]=\text{Im}[E_n(-k)],
\end{gathered}\label{spectrum2}
\end{equation}
such that the bulk currents $I^{+}_{\text{bulk}}$ and $I^{-}_{\text{bulk}}$
cancel each other. On the other hand, the eigenenergies of the chiral edge 
states are real ($\text{Im}[\epsilon(k)]=0$) \cite{supp} so that the electrons  
in chiral edge states are conserved, and the conductance is always quantized.
\par

For unbalanced non-Hermicities, there exists an anomalous CI phase whose 
bulk eigenenergies also satisfy Eqs.~\eqref{spectrum1} and \eqref{spectrum2}, 
as shown in Figs.~\ref{fig3}(e,f). Thus, the bulk states do not contribute 
any net current. Different from the balanced non-Hermicity case, 
the eigenenegies of the edge chiral states are not real any more. 
The imaginary part $\text{Im}[\epsilon(k)]=\gamma$ destroys the electron 
conservation and leads to a non-quantized conductance that 
decays exponentially with $L$ and $\gamma$ \cite{supp}:
\begin{equation}
\begin{gathered}
g=\exp[-2\gamma L/(ta)].
\end{gathered}\label{spectrum3}
\end{equation}
Indeed, our analytical formula Eq.~\eqref{spectrum3} accords 
very well with the simulation results without any fitting 
parameter, see Fig.~\ref{fig2}. 
\par

To understand zero $\Delta g$, we recall origin of the universal 
conductance fluctuation in ordinary disordered metals governed by the 
Hermitian Hamiltonian. The fluctuation is from the random opening and 
closing of a conducting channel as disorder configurations vary due to 
inevitable diffusion of defects and impurities at the finite temperature. 
Each conducting channel contributes exactly one quantum conductance of 
$e^2/h$ no matter what its dispersion relation is in the channel 
\cite{note}. Thus there is no conductance fluctuation in 
the one-dimensional conduction channel, and this explains conductance plateaus 
in quantum Hall systems as well as quantum point contacts in 
ballistic region. Our results suggest that zero conductance 
fluctuation is also true for non-Hermitian conducting channels. 
Disorders may modify electron dispersion relation in a channel, but will 
not introduce an uncertainty into the conductance even though the 
conductance is not quantized, as long as the number of edge channels 
(1 here) does not change in the presence of moderate disorder [$W<W_c$ 
as shown in Fig.~\ref{fig1}(f)]. Interestingly, the conductance will 
eventually go to zero as $W$ increases further beyond $W_c$, 
indicating the anomalous CI will be destroyed by strong disorders. 
The critical disorder $W_c$ can be determined by applying the 
self-consistent Born approximation to the non-Hermitian CI \cite{supp}. 
\par

We would like to make a few remarks before the conclusion.
(1) The edge state conductance and conductance fluctuation reported here for both 
balanced and unbalanced non-Hermicities should be generically applicable 
to edge states in non-Hermitian topological semimetals. (2) In this study, 
the randomness is introduced through the Hermitian on-site potential. 
If the randomness is on the non-Hermicity coefficient $\gamma$, we observe 
that the conductance fluctuation would not be zero any more in the 
anomalous CI phase \cite{supp}, and the non-quantized conductance plateau 
would not be perfect as observed here. 
\par

In conclusion, it is shown that the conductance of chiral edges 
states of non-Hermitian Chern insulators can be quantized (to the 
integer of $e^2/h$) or non-quantized, depending on whether  
the non-Hermicities are balanced or unbalanced.  
However, the conductance in both cases is insensitive to Hermitian 
disorder potential, leading to the zero conductance fluctuation and 
conductance plateaus against disorders and the sample width. 
The conductance plateaus can be destroyed by strong disorders above a 
critical value $W_c$ through Anderson localization. 
The non-quantization of conductance is due to the finite lifetime
of chiral edge states in non-Hermitian CIs.
\par    

\begin{acknowledgments}
This work is supported by the National Natural 
Science Foundation of China (Grants No.~11374249 
and 11704061) and Hong Kong RGC (Grants No.~16301518 and 
16300117). C.W. is supported by UESTC and the China Postdoctoral 
Science Foundation (Grants No.~2017M610595 and 2017T100684).

\par

\end{acknowledgments}


\begin{thebibliography}{99}
\bibitem{gammax} At this stage, we set $\gamma_x=0$ to 
avoid the non-Hermitian skin effect in the longitudinal 
(here is $x$) direction of the two-terminal configuration.
\bibitem{topology} According to the non-Hermitian
topological band theory, the Dirac mass is 
renormalized as $m-|\gamma|^2/2$ \cite{Yao2}. 
Thus, for the parameters considered here
($m=t,\gamma=10^{-3}t$), our system is still
topological nontrivial with chiral edge channels.
\bibitem{note} At zero temperature, the conductance 
of a metal is proportional to the product of 
electron group velocity and density of state 
at the Fermi level. For 1D channels, such as the 
edge states in quantum Hall systems, the density 
of state is proportional to the inverse of the 
group velocity. Thus the conductance of a 
1D channel does not depend on electron 
dispersion relation.
\end{thebibliography}
\end{document}